\documentclass[a4paper]{jpconf}
\usepackage{graphicx}
\usepackage{citesort}
\usepackage{textgreek}

 \def\Mso{\,{\rm M}_\odot}

 \def\gcm{\,{\rm g}\,{\rm cm}^{-3}}
 \def\kms{\, {\rm km}\, {\rm s}^{-1}}
 
 \def\simle{\mathrel{\hbox{\rlap{\hbox{\lower4pt\hbox{$\sim$}}}\hbox{$<$}}}}
 \def\simgr{\mathrel{\hbox{\rlap{\hbox{\lower4pt\hbox{$\sim$}}}\hbox{$>$}}}}
 \def\vw{\, v_{\rm w}}
 \def\Mdot{\, \dot{M}}
 \def\vstar{\, v_\star}
 \def\Msoy{\, \Mso~{\rm yr}^{-1}}
 \def\muG{\,\textmu G}

\def\rD{\, R_{\rm D}}

\def\micron{\, \textmu m}
\def\vinf{\, v_\infty}

\begin{document}
\title{Using numerical models of bow shocks to investigate the circumstellar medium of massive stars}

\author{A.J. van Marle$^1$, L. Decin$^1$, N.L.J. Cox$^1$, Z. Meliani$^{2.3}$}

\address{$^1$Institute of Astronomy, KU Leuven, Celestijnenlaan 200D, B-3001, Leueven, Belgium}
\address{$^2$Observatoire de Paris, 5 place Jules Janssen 92195 Meudon, France} 
\address{$^3$ APC, Universit{\'e} Paris Diderot, 10 rue Alice Domon et L{\'e}onie Duquet, 75205 Paris Cedex 13, France}

\ead{allardjan.vanmarle@ster.kuleuven.be}

\begin{abstract}
Many  massive stars travel through the interstellar medium at supersonic speeds. 
As a result they form bow shocks at the interface between the stellar wind. 
We use numerical hydrodynamics to reproduce such bow shocks numerically, creating models that 
can be compared to observations. 
In this paper we discuss the influence of two physical phenomena, interstellar magnetic fields 
and the presence of interstellar dust grains on the observable shape of the bow shocks of massive stars. 

We find that the interstellar magnetic field, though too weak to restrict the general shape of the bow shock,  
reduces the size of the instabilities that would otherwise be observed in the bow shock of a red supergiant. 
The interstellar dust grains, due to their inertia can penetrate deep into the bow shock structure of a main sequence O-supergiant, 
crossing over from the ISM into the stellar wind. 
Therefore, the dust distribution may not always reflect the morphology of the gas. 
This is an important consideration for infrared observations, which are dominated by dust emission. 

Our models clearly show, that the bow shocks of massive stars are useful diagnostic tools that can used 
to investigate the properties of both the stellar wind as well as the interstellar medium.
\end{abstract}

\section{Introduction}
When a star moves through the interstellar medium (ISM), its wind collides with the interstellar gas. 
If the motion of the star is supersonic with respect to the sound speed in the ISM, this collision leads to the formation of a 
\emph{bow shock} in the region ahead of the star. 
Because the size and shape of a bow shock is determined by the balance between the two ram pressure (of the stellar wind on the inside 
and of the motion of the ISM relative to the star on the outside) stellar wind bow shocks 
are powerful diagnostic tools that can help us determine the properties of the ISM, and the stellar wind. 
In addition, the instabilities that can occur in the bow shock can help us analyse the physical process that are taking place.

\subsection{The general shape of stellar-wind bow shocks}
Assuming that the interaction is supersonic with respect to both the stellar wind and the ISM (which is generally the case), 
the bow shock structure consists of four layers: the free-streaming stellar wind, the shocked stellar wind, the shocked ISM and the unshocked ISM. 
The free-streaming wind is separated from the shocked wind by the wind-termination shock. 
Similarly, the forward shock separates the shocked ISM from the unshocked ISM. 
Between the shocked wind and the shocked ISM lies a contact discontinuity.

The stand-off distance ($\rD$) between the star and the bow shock is determined by the ram pressure of  
the stellar wind and the ISM. 
These are in balance at a distance of 
\begin{equation}
\rD\,=\, \sqrt{\frac{\Mdot \vw}{4\pi\rho_{\rm ISM} \vstar^2}}, 
\label{eq:rd}
\end{equation}
with $\Mdot$ and $\vw$ the mass loss rate and velocity of the stellar wind, $\rho_{\rm ISM}$ the density of the ISM and $\vstar$ 
the velocity of the star with respect to the local ISM \cite{Wilkin:1996}. 
Because $\rD$ denotes the distance at which the stellar wind and the ISM are in equilibrium with one-another it actually gives us the location of the 
contact discontinuity, rather than either the wind termination shock, or the forward shock. 
In addition, we can use analytical approximations to determine the opening angle of the bow shock, which \cite{Wilkin:1996} 
described as
\begin{equation}
 \frac{R(\theta)}{\rD}\,=\,\frac{1}{\sin{\theta}} \sqrt{3\biggl(1-\frac{\theta}{\tan{\theta}}\biggr)}
\end{equation}
with $\theta$ the angle between the direction of motion of the star and a line from the star to a particular point along the bow shock. 

\section{Numerical method} 
\label{sec-num}
Although it is possible to predict the morphology of a bow shock analytically, 
including the general 2-D structure (E.g.~\cite{Dganietal:1996,SchulreichBreidschwerdt:2011}), 
such analytical models are inherently limited in the amount and type of physical processes 
they can include as well as in their inability to quantitatively reproduce instabilities. 
Therefore, it becomes necessary to use numerical simulations to model the stellar wind bow shocks. 
We use the {\tt MPI-AMRVAC} magneto-hydrodynamics code \cite{vanderHolstetal:2008,Keppensetal:2012}, which solves 
the conservation equations for mass, momentum and energy on an adaptive mesh grid. 
For our calculations we include radiative cooling using the method described in \cite{vanMarleKeppens:2011}, with a cooling curve 
for solar metallicity.

We model each bow shock in the co-moving frame of the star.
We start our simulations by filling a 2-D cylindrical grid in the R,Z-plane with a constant density interstellar medium, 
which has a constant velocity parallel to the Z-axis and is kept constant by allowing material to flow in at the outer Z-boundary.  
At the lower Z-boundary, the material is allowed to flow out of the grid. 
The stellar wind is introduced by filling a small half sphere, centred on the origin, with gas according to a free-streaming stellar wind profile 
(constant velocity and density decreases with the radius squared).

\begin{table}[htp]
\centering
 \caption{Physical parameters of the wind of \textalpha-Orionis and the local ISM, used as input for our simulations. }
\label{Table:input_AOri}
\begin{tabular}{lccl}
 \hline \hline
 \noalign{\smallskip}
Mass loss rate                  & $\Mdot$\,&=&\,$3.0\times10^{-6}$\,$\Msoy$ \\
Wind velocity       & $\vinf$\,&=&\,15.0\,$\kms$\\
Velocity w.r.t.\ ISM    & $v_\star$\,&=&\,28.3\,$\kms$\\
ISM density            & $\rho_{\rm ISM}$\,&=&\,$10^{-23.5}\gcm$ \\
ISM temperature           & $T_{\rm ISM}$\,&=&\,10\,K \\
ISM magnetic field                  & $B$\,&=& 3.0 \muG\, \\
\noalign{\smallskip}
\hline
\end{tabular}\\
\end{table}

\section{The influence of interstellar magnetic fields on the bow shock of \textalpha-Orionis}
\subsection{Background}
Recent \emph{Herschel} observations \cite{Decinetal:2012} show us that the bow shock of \textalpha-Orionis, 
a red supergiant (RSG) type evolved star is smooth, without large instabilities. 
However, both analytical predictions \cite{Dganietal:1996} and numerical models \cite{BrighentiDercole:1995,Comeronkaper:1998,vanMarleetal:2011} 
indicate that a bow shock of this kind, where the stellar velocity through the ISM is larger than the wind velocity, 
should show large scale instabilities. 
Several explanations for this discrepancy have been offered, 
such as the possibility that the bow shock is too young \cite{Mohamedetal:2012,Mackeyetal:2012}, 
or that the presence of ionizing photons reduces the instability \cite{Meyeretal:2014}. 
An alternative explanation, as shown in \cite{vanMarleetal:2014}, is that the interstellar magnetic field 
inhibits the growth of instabilities.

\subsection{Interstellar magnetic fields}
The ISM in the galaxy contains magnetic fields than can stretch out over large distances 
($\simeq\,100$\,pc) \cite{RandKulkarni:1989,OhnoShibata:1993,Beck:2009,Shabalaetal:2010}.
Estimates for the magnetic field in the Orion~arm of the Galaxy at a distance of 8\,000\,kpc from the Galactic centre 
(corresponding to the approximate location of \textalpha-Orionis) 
range from $1.4\pm0.3$\muG\, \cite{Fricketal:2001} through 2-3\muG\, in the region near our solar system \cite{HeerikhuisenPogorelov:2011} 
to 3.7-5.5\muG\, as obtained from \emph{Voyager} measurements \cite{Opheretal:2009}. 
These values coincide with the values for the interstellar field in the galaxy at large obtained from \emph{WMAP} data \cite{JanssonFarrar:2012a,JanssonFarrar:2012b}. 

\subsection{Numerical setup}
We use the basic numerical set-up described in Section~\ref{sec-num}, but include a magnetic field parallel to the Z-axis (and therefore parallel 
to the direction of motion of the star). 
This direction is chosen to preserve the 2-D symmetry of the problem and avoid the necessity to run the simulation in 3-D. 
For our computational grid we choose an initial size of of 160\,$\times$\,160\,grid cells, covering a physical domain 
of 2\,$\times$\,2\,pc. 
Using the adaptive mesh option of {\tt{MPI-AMRVAC}}, we allow the code to refine this grid up to four times, depending on variations in the local gas velocity, 
which gives us 2\,560\,$\times$\,2\,560 grid cells effectively. 
All physical input parameters for our model are give in Table~\ref{Table:input_AOri} and are based on observational data from \cite{Uetaetal:2008}.
We run these simulations twice, once with and once without the magnetic field in order to compare the results.

\subsection{Results}
\begin{figure}[h]
\centering
\begin{minipage}{17pc}
\includegraphics[width=17pc]{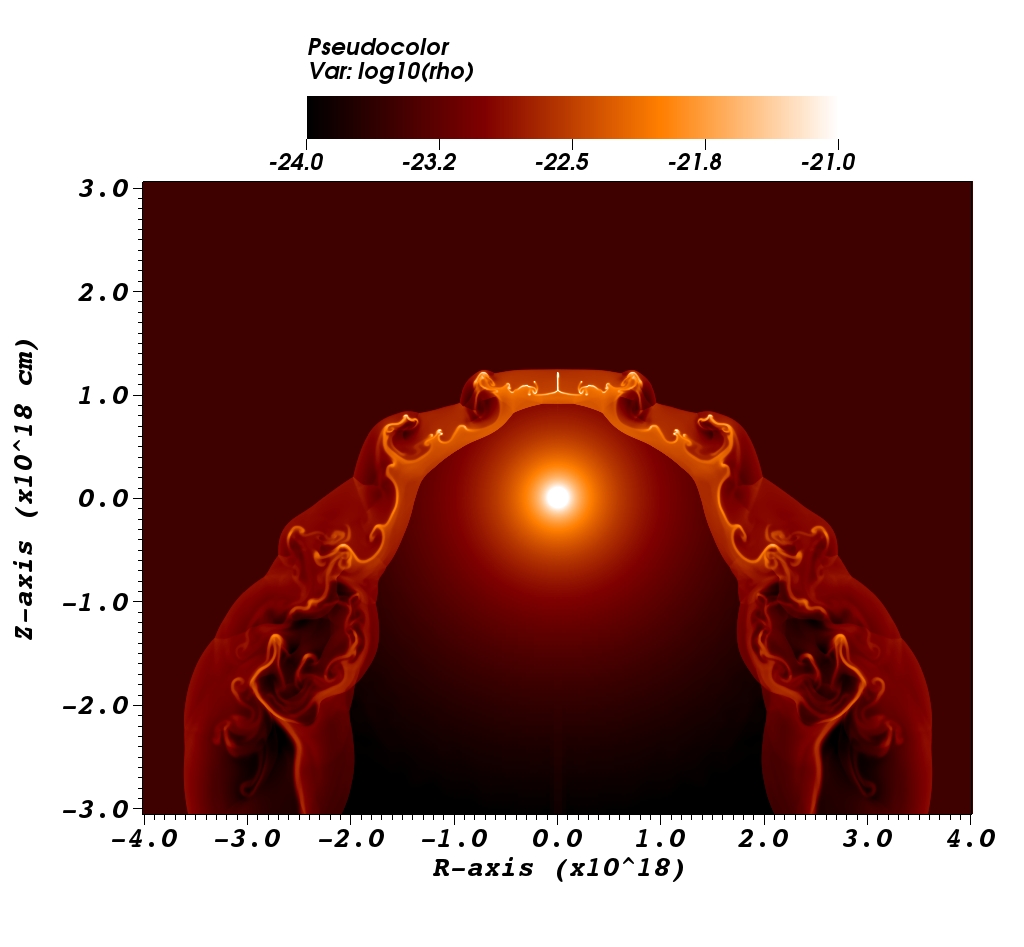}
\caption{\label{fig:Aori_B00} Density in $\gcm$ for the non-magnetic model of the bow shock of \textalpha-Orionis. 
Note that the 2-D model has been reflected in the Z-axis to show a complete bow shock. 
The interface between the shocked wind and shocked ISM is clearly unstable.}
\end{minipage}\hspace{2pc}%
\begin{minipage}{17pc}
\includegraphics[width=17pc]{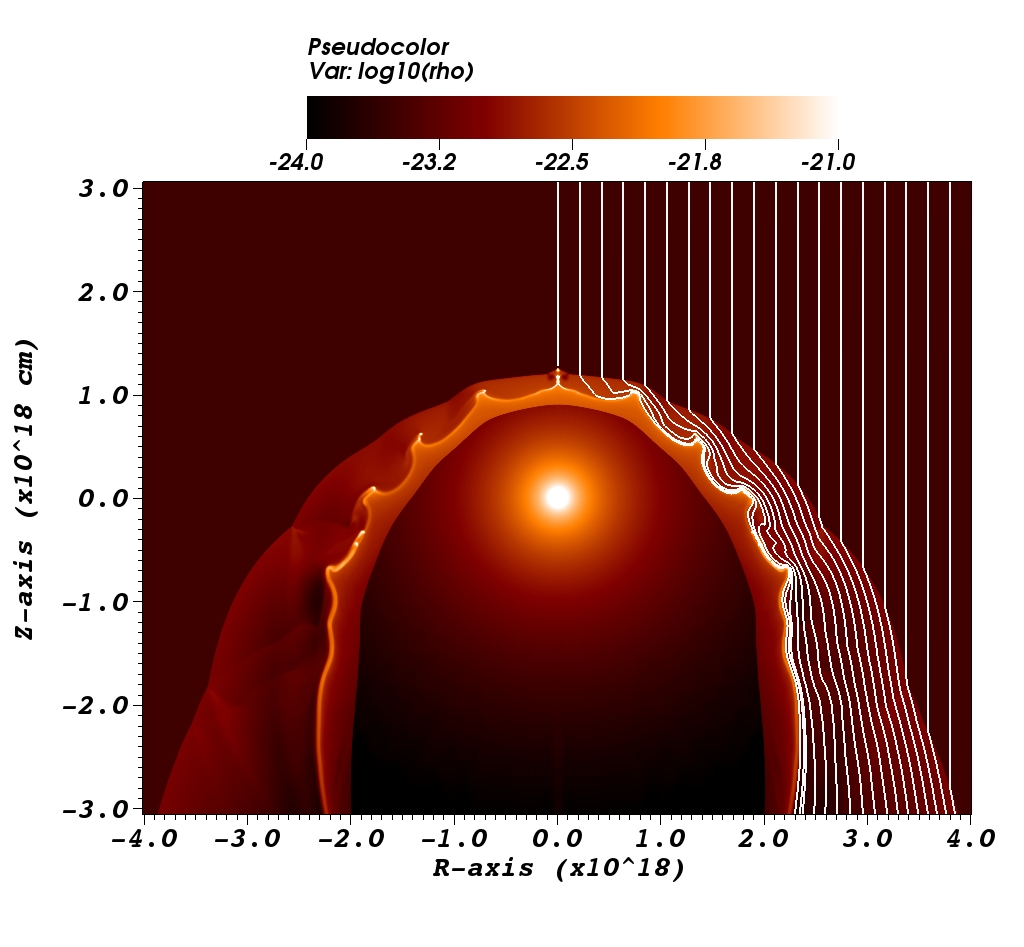}
\caption{\label{fig:Aori_B03} Similar to Fig.~\ref{fig:Aori_B00}, but with an interstellar magnetic field of 3\micron. 
The field lines of the magnetic field are shown in the right side of the panel. 
The instabilities are much smaller than for the non-magnetic model.}
\end{minipage} 
\end{figure}

The result of our simulations, shown in Figs.~\ref{fig:Aori_B00} and \ref{fig:Aori_B03}, clearly demonstrate the influence of the magnetic field. 
Figure~\ref{fig:Aori_B00} shows the model without a magnetic field after 100\,000 years physical time. 
The bow shock is clearly unstable at the contact discontinuity with both Kelvin-Helmholtz and Rayleigh-Taylor instabilities. 
These instabilities, which start out small in the region directly ahead of the star increase in size when they move down stream 
until they grow so large that they distort the general shape of the bow shock. 
For the model that includes an interstellar magnetic field (Fig.~\ref{fig:Aori_B03}, the result is completely different. 
Instabilities do form initially at the front of the bow shock. 
However, rather than growing in size, they remain small as they move downstream. 
Also, the small scale structures that are clearly visible in the non-magnetic model are completely absent 
and the instabilities are limited to a single wavelength. 
This behaviour corresponds to the predictions of, for example, \cite{Junetal:1995,Breitschwerdtetal:2000}, which show that 
a magnetic field inhibits the growth of instabilities with short wavelengths, with the critical wavelength determined by the magnetic field strength, 
the density contrast across the discontinuity and the angle between the wave-vector and the magnetic field. 
This was further demonstrated by \cite{vanMarleetal:2014}, which showed models for a range of magnetic field strengths. 

\subsection{Discussion}
It is clear from our models that the presence of an interstellar magnetic field can suppress the growth 
of instabilities in the bow shock of a RSG type star. 
Of course, this model is limited by the fact that we have to align the magnetic field with the direction of motion of the star. 
Further study (in 3D) will be necessary to quantify this effect for magnetic fields at other angles. 

\begin{table}[htp]
\centering
 \caption{Physical parameters of the wind of an O-supergiant and the local ISM, used as input for our simulations. }
\label{Table:input_Ostar}
\begin{tabular}{lccl}
 \hline \hline
 \noalign{\smallskip}
Mass loss rate                  & $\Mdot$\,&=&\,$1.0\times10^{-5}$\,$\Msoy$ \\
Wind velocity       & $\vinf$\,&=&\,2300\,$\kms$\\
Velocity w.r.t.\ ISM    & $v_\star$\,&=&\,77\,$\kms$\\
ISM density            & $\rho_{\rm ISM}$\,&=&\,$10^{-23.5}\gcm$ \\
ISM temperature           & $T_{\rm ISM}$\,&=&\,10\,000\,K \\
dust grain sizes          & $a_1, a_2, a_3$\,&=&\, 0.071\micron, 0.19\micron, 0.37\micron  \\        
\noalign{\smallskip}
\hline
\end{tabular}\\
\end{table}

\section{Dust and gas in circumstellar bow shocks}
\subsection{Background}
The introduction of  satellites like {\emph{Spitzer}} and {\emph{Herschel}}, has allowed us to resolve circumstellar structures,such as bow shocks, in the infrared. 
However, rather than observing the morphology of the gas directly, these infared observations 
actually show us the distribution of dust grains, which are the primary source of infrared radiation. 
Therefore, it is absolutely necessary to investigate whether such dust grains, typically more than 0.5\% of the total circumstellar mass, 
are representative for the gas. 
This was done for a red supergiant type star by \cite{vanMarleetal:2011} where the star itself was assumed to be the primary source of dust. 
These simulations showed that larger dust grains ($>\,0.045$\micron) tend to decouple from the gas, once the gas is 
decelerated by the bow shock.

We now investigate the behaviour of dust grains in the bow shock of a hot star, where the situation is reversed. 
The wind of a hot massive star does not contain dust, therefore the main source for dust is the interstellar medium.

\subsection{Numerical setup}
We use the basic numerical set-up described in Section~\ref{sec-num} and include the 
presence of the interstellar dust grain by filling the interstellar medium with dust grains of three different radii 
(radii of 0.071\micron, 0.19\micron, and 0.37\micron respectively), each representing a 'bin' 
of grain sizes. 
These dust grains are treated as pressure-less gasses according to the same method described in \cite{vanMarleetal:2011}, 
with the interaction between dust and gas included in the form of a drag force \cite{Kwok:1975}. 
We assume that the total dust mass equals 0.5\% of the ISM gas mass. 
The number densities of the three dust types are scaled in such a way that al three 'bins' contain an equal amount of mass and 
that the dust follows the size distribution of $n(a)\,\propto\,a^{-3.5}$ with 
$n$ the particle density and $a$ the grain radius described by \cite{DraineLee:1984}. 
The stellar wind and ISM parameters for our simulation are given in Table~\ref{Table:input_Ostar}, 
based on the O4~supergiant~BD+43\,3654 \cite{ComeronPasqali}, 
with stellar wind properties for such a star estimated according to \cite{Muijresetal:2012}. 
We assume a warm ISM (10\,000\,K) because the stellar radiation can be expected to ionize the surrounding hydrogen, 
creating an HII region that extends well beyond the bow shock \cite{McKeeetal:1984}.

For this simulation we use a basic grid of $160\times160$ cells covering a physical domain of $10\times10$ parsec. 
The adaptive mesh is allowed four additional levels, giving us a maximum effective grid of $2\,560\times2\,560$ cells.

\subsection{Results}
\begin{figure}[h]
\centering
\begin{minipage}{17pc}
\includegraphics[width=17pc]{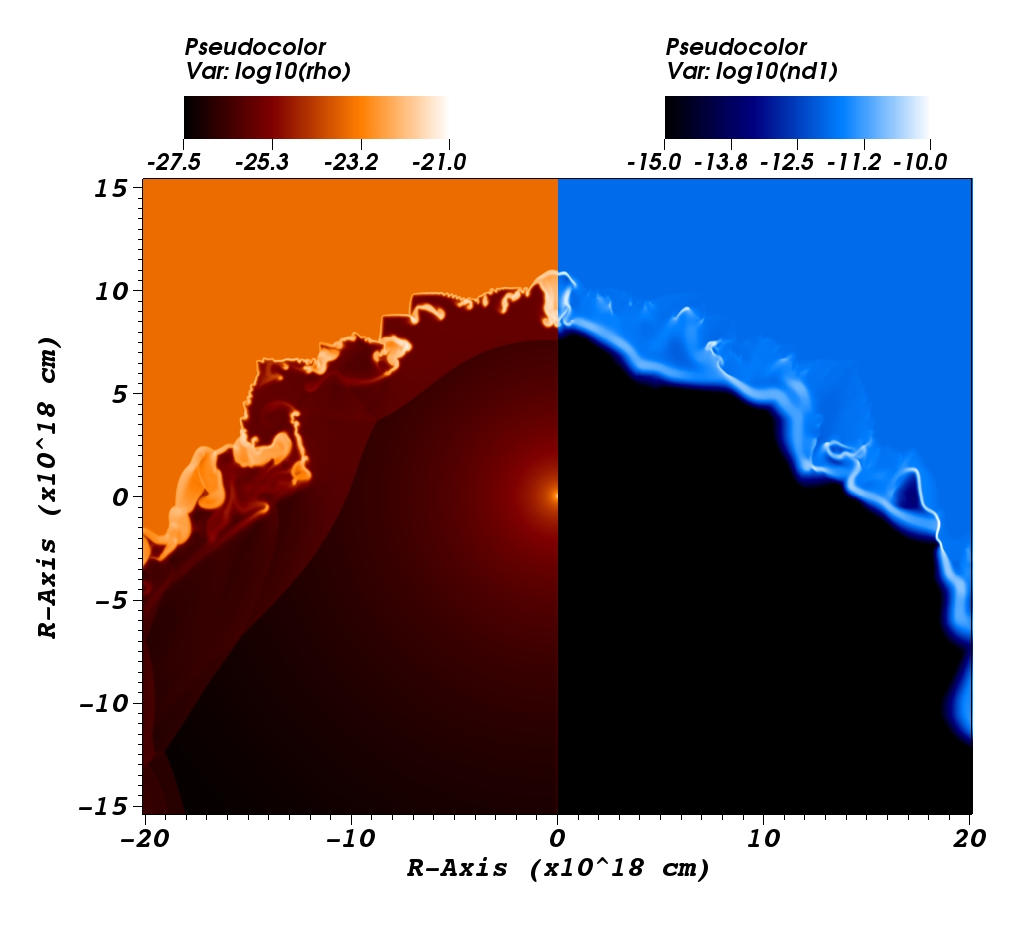}
\caption{\label{fig:Ostar_gas}Gas density in $\gcm$ of the gas (left) and number density for small dust grains (right) 
for the bow shock of a massive O4 supergiant. The gas shows a thin, unstable shell of shocked ISM, 
whereas the dust grains are spread out over 
the entire shocked gas region with the highest concentration at the wind termination shock. }
\end{minipage}\hspace{2pc}%
\begin{minipage}{17pc}
\includegraphics[width=17pc]{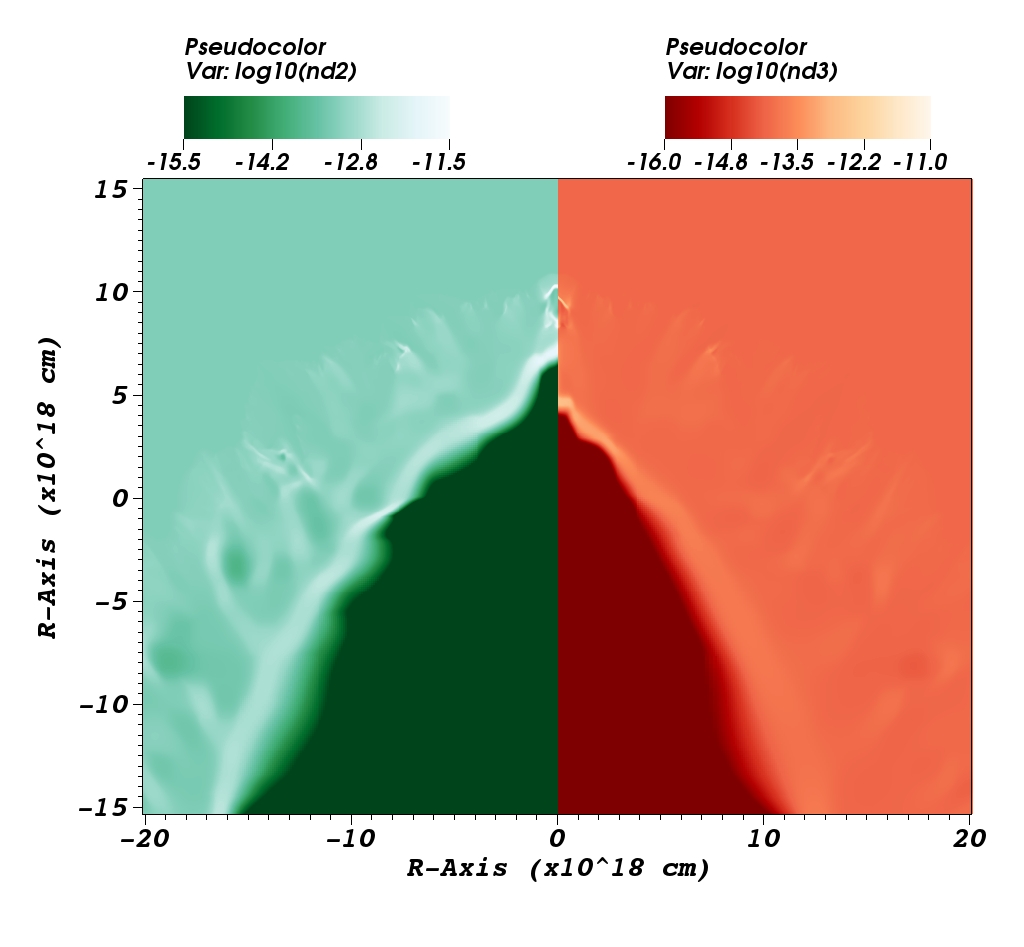}
\caption{\label{fig:Ostar_dust} Similar to Fig.~\ref{fig:Ostar_gas}, but for the number density of 
 intermediate (left) and large (right) dust grains. 
 Owing to their larger inertia, these grains cross both the shocked ISM and the shocked wind and enter the free-streaming wind region 
 before the drag-force stops them, with the largest grains showing the deepest penetration.}
\end{minipage} 
\end{figure}

The result of our simulation is shown in Figs.~\ref{fig:Ostar_gas} and\ref{fig:Ostar_dust}. 
The gas (left side of Fig.~\ref{fig:Ostar_gas}) shows that the forward shock is highly radiative, causing the shocked ISM to be 
compressed into a thin shell. 
The smallest dust grains (right side of Fig.~\ref{fig:Ostar_gas}) show much less compression. 
These grains, which originate in the ISM, penetrate the shocked ISM shell and enter the shocked wind region behind it. 
They only come to a stop at the wind termination shock due to the increases drag force 
generated by the unshocked wind moving in the opposite direction. 
the intermediate and large dust grains (left and right side of Fig.~\ref{fig:Ostar_dust}) show that these grains, 
which have a larger momentum compared to their surface area are able to penetrate into the unshocked wind. 

\subsection{Discussion}
Our simulations show that under the circumstances of our particular model, the dust distribution can deviate considerably from 
the gas morphology. 
This can have serious consequences for observations of such bow shocks. 
In this particular case, observations at visual wavelengths, dominated by gas emission 
would show a thin, highly compressed bow shock (the high density shell of shocked ISM), 
whereas infrared observations, dominated by the dust, would show a thick shell.

\section{Conclusions}
We have shown that numerical models can be used to reproduce the bow shocks of massive stars. 
These models can also be used to investigate the effect of physical phenomena, such as magnetic fields and the presence of dust grains, 
which can influence the structure of the bow shock and/or the manner in which it will appear in observations. 
In the future we hope to continue our research in this field by including additional physical effects, such as 
thermal conduction and the interaction between dust grains and the magnetic field. 
We also intend to extend our models to 3-D in order to investigate what occurs/ when the 2-D symmetry is broken. 
E.g. when magnetic field and direction of motion are not aligned.

\ack 
A.J.v.M.\ acknowledges support from FWO, grant G.0277.08, K.U.Leuven GOA/2008/04 and GOA/2009/09.

\section*{References}
\bibliography{vanMarle_13thannual}
 {\typeout{}
  \typeout{******************************************}
  \typeout{** Please run "bibtex \jobname" to obtain}
  \typeout{** the bibliography and then re-run LaTeX}
  \typeout{** twice to fix the references!}
  \typeout{******************************************}
  \typeout{}
 }

\end{document}